\documentclass[aps,prd,showpacs,preprintnumbers]{revtex4}
\usepackage{graphics}
\def \beq{\begin{equation}}
\def \eeq{\end{equation}}
\def \beqa{\begin{eqnarray}}
\def \eeqa{\end{eqnarray}}
\def \N{{\cal N}}
\def \Q{{\cal Q}}
\def \etc{{\sl etc.\/}}

\def \ie{{\sl i.e.\/}}
\def \etal{{\sl et al.\/}}

\def \jp{{\sl J.\ Phys.\ }}

\def \pl{{\sl Phys.\ Lett.\ }}
\def \pr{{\sl Phys.\ Rev.\ }}
\def \prl{{\sl Phys.\ Rev.\ Lett.\ }}

\begin{document}
 
\title{Fluctuations, strangeness and quasi-quarks\\
   in heavy-ion collisions from lattice QCD}
\author{R.\ V.\ \surname{Gavai}}
\email{gavai@tifr.res.in}
\affiliation{Department of Theoretical Physics, Tata Institute of Fundamental
         Research,\\ Homi Bhabha Road, Mumbai 400005, India.}
\author{Sourendu \surname{Gupta}}
\email{sgupta@tifr.res.in}
\affiliation{Department of Theoretical Physics, Tata Institute of Fundamental
         Research,\\ Homi Bhabha Road, Mumbai 400005, India.}

\begin{abstract}
We report measurements of diagonal susceptibilities for the baryon
number, $\chi_B$, electrical charge, $\chi_Q$, third component of
isospin, $\chi_I$, strangeness, $\chi_S$, and hypercharge, $\chi_Y$,
as well as the off-diagonal $\chi_{BQ}$, $\chi_{BY}$, $\chi_{BS}$,
\etc. We show that the ratios of susceptibilities in the high
temperature phase are robust variables, independent of lattice
spacing, and therefore give predictions for experiments. We also
investigate strangeness production and flavour symmetry breaking
matrix elements at finite temperature. Finally, we present evidence
that in the high temperature phase of QCD the different flavour
quantum numbers are excited in linkages which are exactly the
same as one expects from quarks. We present some investigations of
these quark-like quasi particles.
\end{abstract}
\preprint{TIFR/TH/05-41}
\maketitle

\section{Introduction}

Experiments plan to study fluctuations of conserved quantities in
heavy-ion collisions at the RHIC and LHC in different rapidity
windows. With proper particle
identification, one can measure in the experiment both absolutely
conserved quantities like the baryon number ($B$) and the electrical
charge ($Q$), as well as quantities which are conserved only under
the strong interactions, such as  the third component of isospin
($I_3$), the strangeness ($S$) and the hypercharge ($Y$). These
observations can be used to extract fluctuations in the numbers of
these quantities \cite{ahm,jk}.  Such observations need to be
compared to predictions of quark number susceptibilities (QNS) from
lattice QCD. In this paper we report on lattice computations of a
variety of diagonal QNS--- $\chi_B$, $\chi_Q$, $\chi_I$, $\chi_S$
and $\chi_Y$. One of the main results in this paper is the extraction
of predictions for the ratios of these susceptibilities which survive
the continuum limit.  Our second important result is the investigation
of the strange quark sector of the theory: we extract the Wroblewski
parameter in a dynamical QCD computation for the first time, and
also investigate the dynamics and kinematics of flavour symmetry
breaking in QCD. Further, we present results on the cross correlations
$\chi_{BQ}$, $\chi_{BY}$, $\chi_{BS}$ and $\chi_{QY}$. These cross
correlations are used to explore the charge and baryon number of
objects that carry flavour. We find that the baryon number of flavour
carrying objects immediately above the QCD crossover temperature,
$T_c$, are 1/3 and the charges are 1/3 or 2/3. We find furthermore,
that these objects are almost pure flavour--- anything carrying u
flavour has only tiny admixtures of d and s flavours, \etc. This
is our third main result.

We have bypassed the necessity of numerically taking the continuum
limit of the theory by restricting attention to the high temperature
phase where it is easy to define robust observables which have
little, or no, lattice spacing dependence.
We demonstrate the robustness of
the observables in quenched QCD, and then compute these quantities
in QCD with two flavours of light dynamical quarks. These are also
good observables in the sense of \cite{ahm}---
\beq
   C_{K/L} \equiv \frac{\chi_K}{\chi_L}
     = \frac{\sigma^2_K}{\sigma^2_L} \,,
\label{ratio}\eeq
where $\chi_K$ and $\chi_L$ are QNS for the conserved quantum numbers
$K$ and $L$ and 
$\sigma_K$ and $\sigma_L$ are the variances. The two variances must be
obtained under
identical experimental conditions, after removing counting (Poisson)
fluctuations as suggested by \cite{pruneau}. Thus the robust lattice
observables give predictions for robust experimental observables.

Either of $K$ or $L$ can also stand for a composite label $(M,N)$
where $M$ and $N$ are conserved quantum numbers--- in this case the
susceptibility is an off-diagonal susceptibility, and the variance
has to be replaced by the covariance of $M$ and $N$. Note the
relation with the correlation coefficient---
\beq
   r_{MN} = \frac{\langle MN\rangle -\langle M\rangle
                     \langle N\rangle}{\sigma_M\sigma_N}
     = \frac{\chi_{MN}}{\sqrt{\chi_M\chi_N}}
     = C_{(M,N)/M}\sqrt{C_{M/N}} = C_{(N,M)/N}\sqrt{C_{N/M}}
\label{robcor}\eeq
where again, the expressions are robust both on the lattice and
in experiment. The study of these robust variables tells us about
the relative magnitudes of fluctuations in different quantum
numbers. The study of these quantities is one of the main results
reported here.

We further present investigations of the strange quark sector of the theory.
The robust variable $C_{SU}$ is closely related to the Wroblewski
parameter which can be extracted from experiments. This shows strong
dependence on the actual strange quark mass, $m_s$, in the vicinity
of $T_c$. Since $m_s\simeq T_c$, it seems that part of this sensitivity
could be attributed purely to kinematics. We investigate the dynamical
matrix elements which are responsible for flavour symmetry breaking in
QCD and compare the importance of kinematics and dynamics in the
strange quark sector. This is our second major result.

One outstanding question about the high temperature phase of QCD
is the nature of flavoured excitations. There is ample evidence
that quarks are liberated at sufficiently high temperature--- the
continuum limit of lattice computations of screening masses are
consistent with the existence of such a Fermi gas for $T\ge2T_c$
\cite{mtc,valence}; quantitative agreement between weak coupling
estimates of the susceptibilities \cite{bir,alexi} and the lattice
data \cite{pushan,valence} also confirm this; the equation of state
at very high temperature also testifies to this. However, comparison
of lattice results and weak coupling computations of these quantities
fail for $T<2T_c$. Our third new result concerns this matter of the
thermodynamically important single particle excitations.

We address this question in the most direct way possible--- create
an excitation with one quantum number and observe what other quantum
numbers it carries. Technically, this involves the measurement of
robust ratios of off-diagonal QNS; the correlation between quantum
numbers $K$ and $L$ can be studied through the ratio
\beq
   C_{(KL)/L} = \frac{\langle KL\rangle-\langle K\rangle\langle L\rangle}{
     \langle L^2\rangle-\langle L\rangle^2}.
\label{example}\eeq
We find that such measurements are feasible on the lattice, and are
open to direct interpretation. We also suggest that they could be
performed in heavy-ion experiments, as direct tests of whether
quarks exist in the hot and dense matter inside the fireball. A
recent suggestion of \cite{koch} is the measurement of just such a
variable: essentially $C_{(BS)/S}$.

We find that, immediately above $T_c$, the baryon number, charge and
other flavour quantum numbers are linked with each other in exactly the
same way as they are in quarks. For example, excitations which carry
unit strangeness carry baryon number of $-1/3$ and charge of $+1/3$.
This, together with the fact that there is also a failure of weak
coupling theory, would imply that the QCD plasma phase is a ``quark
liquid'' in the sense that the quasi-particles carry the quantum
numbers of quarks, but the interactions between them are too strong for
the system to be treated in weak coupling theory.  Extension of these
measurements to finite chemical potential for $T>T_c$ and $\mu\gg T$
could allow us to check whether or not the system is a normal Fermi
liquid \cite{landau}. Such an extension is feasible since the Taylor
series expansion of the free energy in $\mu/T$ has a radius of
convergence much higher than unity for $T>T_c$.

This is an appropriate place to remark upon a few aspects of our
computations.  Having removed most of the lattice spacing uncertainties
by using robust variables, we have to control only the quark masses.
We do this partly by performing the computations in an approximation
called partial quenching. In this approximation the valence quark
masses in the theory are tuned keeping the sea quark masses fixed.
We explore the dependence of the robust variables on the sea quark
masses and find that the results are not very sensitive to these
parameters. This is expected--- away from a phase transition there
is no more than a 5\% change in the QNS in going from quenched to
$N_f=2$ dynamical QCD, and one expects the change to be smaller in
going from $N_f=2$ to $N_f=2+1$, as long as one avoids the vicinity
of the phase transition.  The ratios are even less sensitive to the
sea quark content than the QNS. In this study we have concentrated
on the numerically more important effect of the valence quark masses.

We have used two flavours of dynamical sea quarks of bare mass
$m=0.1T_c$ to study a temperature range upto about
$2T_c$.  These quark masses are such that $m_\rho/T_c=5.4$ and
$m_\pi/m_\rho=0.3$--- which makes this the smallest quark mass used
in a systematic study of fluctuations.  We have taken the strange
quark to be quenched and to have a bare mass in the range
$m_s/T_c=0.75$--1. This gives the correct physical values of the
ratio $m_K/m_\rho$. We have also investigated the effect of decreasing
the valence light quark mass by a factor of three in order to get
at the same time the correct physical value of the ratio $m_\pi/m_\rho$,
and varying the strange quark mass about the physical value.

Details of simulations and the results are given in the next section,
and a summary of the results in the final section. Details of the
formalism, including expressions for various QNS are given in the
appendix.

\section{Simulations and results}

\subsection{The simulations}

In earlier papers \cite{first,endpt,nls} we have shown that finite
volume effects on the QNS are negligible for lattices with $N_s\ge2N_t$
($N_s$ is the spatial extent of the lattice and $N_t$ the temporal
extent). The data we discuss here are obtained on $4\times16^3$
lattices. The setting of scale, the parameters employed and the statistics
are detailed in \cite{endpt}. To that set of data 
we have added two more sets--- 55 configurations separated by more than two
autocorrelation times at $T/T_c=0.975\pm0.010$ (\ie, $\beta=5.2825$)
and 86 configurations, similarly spaced, at $T/T_c=1.15\pm0.01$
(\ie, $\beta=5.325$). The configurations are generated with a bare
sea quark mass $m=0.1T_c$, which gives $m_\pi=0.3 m_\rho$.

We have explored the dependence of the physics on the strange quark
mass and on variations in the light quark mass through partially
quenched computations, \ie, the approximation in which the number
of valence quark flavours is different from the number of dynamical
sea quark flavours, and their masses are also different.  Errors
in partial quenching are bounded by comparing results with the fully
quenched theory.

\subsection{Quark number susceptibilities}

\begin{figure}
\begin{center}
   \scalebox{0.7}{\includegraphics{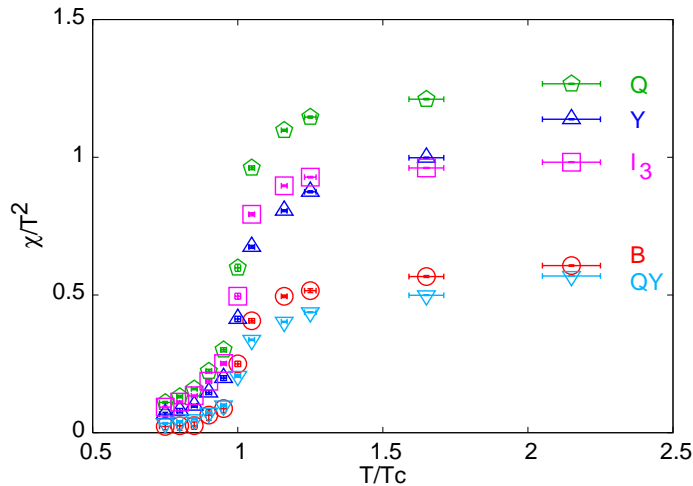}}
\end{center}
\caption{Some of the QNS, $\chi/T^2$, as functions of $T/T_c$ for
   $m_{ud}=0.1T_c$ and $m_s=T_c$.}
\label{fg.qns}\end{figure}

Our primary results for QNS are shown in Figure \ref{fg.qns}. These
were obtained using the eqs. (\ref{qnsa}) and (\ref{qnsb}) in Appendix
\ref{sc.qns}. The diagonal QNS and several of the off-diagonal ones show the
characteristic crossover from small values in the low temperature
phase to large values in the high temperature phase which gave rise
to the original interpretation that the QCD phase transition liberates
quarks \cite{milc,gavai}.  Observe that $\chi_B < \chi_Q$ through the
full temperature range explored.  Both $\chi_I$ and $\chi_Y$ have
values between the two others.  In the low temperature phase one
has $\chi_Y < \chi_I$, but for $T\ge1.5 T_c$ one obtains $\chi_Y >
\chi_I$. We expect the crossover temperature between these two
regimes to vary with quark masses. 

Our results are compatible with earlier results with staggered
fermions at the same cutoff and quark mass which were obtained in
the high temperature phase \cite{pushan}. They are not directly
comparable to results obtained in \cite{milc2} at the same lattice
spacing due to differences in the discretization.

\subsubsection{Robust observables}

\begin{figure}
\begin{center}
   \scalebox{0.5}{\includegraphics{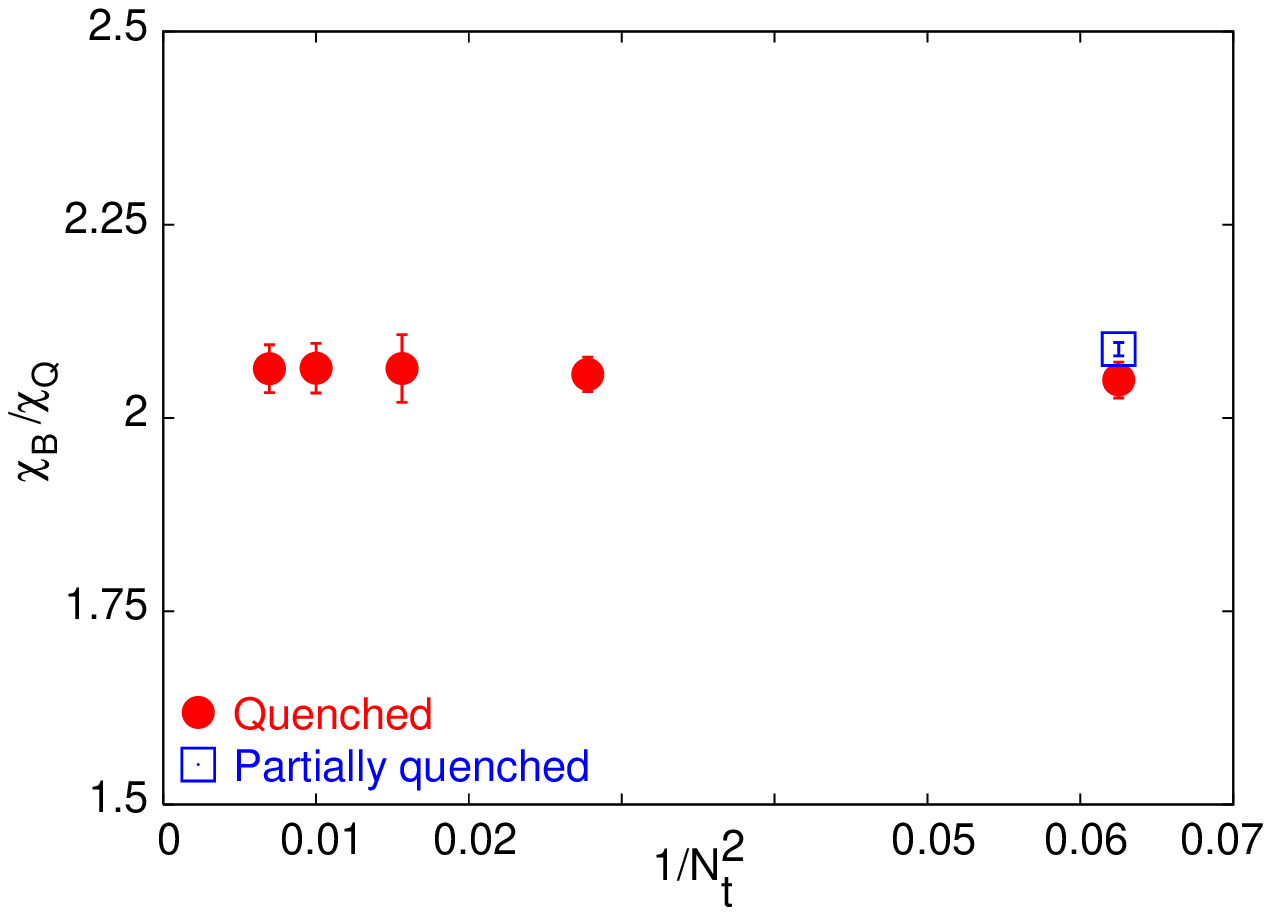}}
   \scalebox{0.5}{\includegraphics{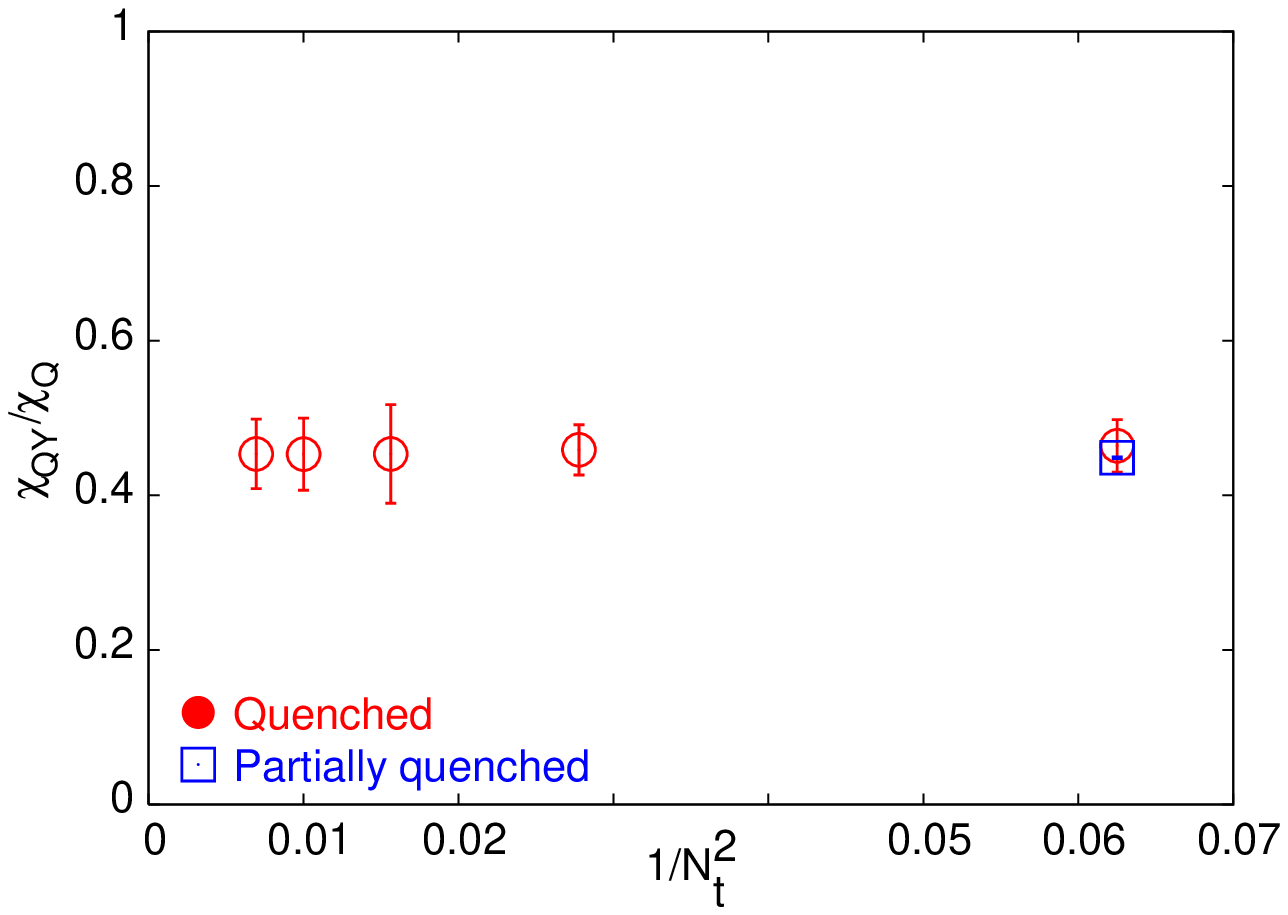}}
\end{center}
\caption{Ratios of QNS are robust observables--- being insensitive 
   to both changes in lattice spacing $a\propto1/N_t^2$ at fixed $T=2T_c$,
   and the sea quark content of QCD. The quenched results come from a
   reanalysis of data from \cite{conti}. In both cases the light valence
   quark mass is $0.03T_c$ and the strange quark mass is $T_c$.}
\label{fg.ratio}\end{figure}

\begin{figure}
\begin{center}
   \scalebox{0.7}{\includegraphics{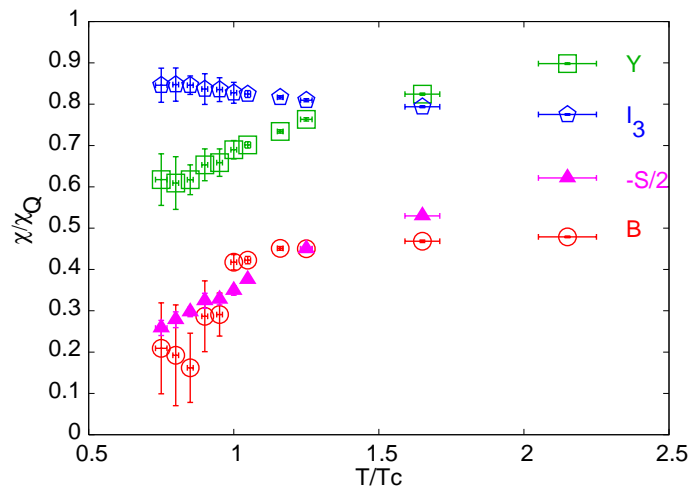}}
\end{center}
\caption{Some robust predictions of fluctuation measures from QCD: all
   the quantities shown are the ratio $C_{X/Q}$ for the $X$ indicated
   in the figure, except for $X=S$ which is $C_{S/Q}/2$.}
\label{fg.robust}\end{figure}

In the quenched theory it was found that the QNS depended quadratically
on the lattice spacing \cite{conti}, \ie, $\chi(a)=\chi+{\cal
O}(a^2)$. Since staggered Fermions have order $a^2$ lattice artifacts,
one expects the same behaviour in the theory with sea quarks. We
are therefore forced to search for observables which are robust
against changes in the lattice spacing, in the sense that $r(a)=r+{\cal
O}(a^n)$ with $n>2$. We expect the ratios of QNS to have very good
scaling properties in the high temperature phase, where the flavour
off-diagonal QNS are much smaller than the flavour diagonal QNS.
In the low-temperature phase we do not necessarily expect such
behaviour to hold, since these two pieces are comparable, and the
coefficient of the order $a^2$ corrections in the two parts depend
on different physical quantities.

As shown in Figure \ref{fg.ratio}, ratios of QNS in the high
temperature phase have this property.  The figure also shows another
pleasant property--- these ratios have little statistically significant
dependence on the sea quark content of the theory. We have checked
that these two aspects of robustness hold for all ratios in the
high temperature phase of QCD. The dependence of such ratios on the
valence quark masses can be determined using the quadratic response
coefficients (QRC) defined in \cite{rajarshi} and applied to the
study of $C_{B/S}$.

In view of these results, the hierarchy of QNS shown in the previous
subsection must be a robust feature of QCD. It is therefore useful
to demonstrate this hierarchy by plotting $C_{X/S}$ as a function
of $T/T_c$ in Figure \ref{fg.robust}. Our results indicate that
experimental studies of $C_{S/Q}$, $C_{B/Q}$ and $C_{Y/Q}$ are the
most promising in terms of distinguishing between the two phases
of QCD, because they exhibit the largest changes in going from one
phase to the other.

\subsection{Strange quarks}

\begin{figure}
\begin{center}
   \scalebox{0.7}{\includegraphics{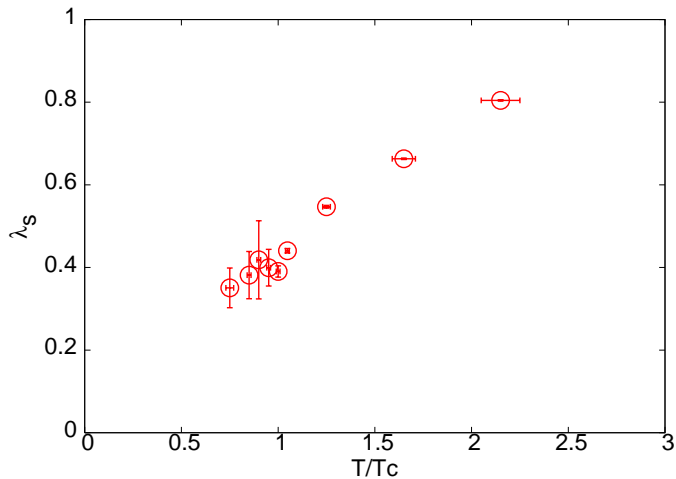}}
\end{center}
\caption{The robust variable $C_{s/u}=\lambda_s$ as a
   function of $T/T_c$ when the light quark masses are taken to be
   $m_{ud}=0.03T_c$, corresponding to a realistic pion mass, and
   the strange quark mass is set to $m_s=T_c$, which gives a realistic
   value of the ratio $m_K/m_\pi$.}
\label{fg.wrob}\end{figure}

The Wroblewski parameter, $\lambda_s$, as extracted from experiments,
is the ratio of the numbers of primary produced strange
and light quark pairs.  It has been argued earlier \cite{valence}
that under certain conditions, whose satisfaction can be verified
by independent observations, one has $\lambda_s=C_{s/u}$.  Our results 
for this robust 
quantity are shown in Figure \ref{fg.wrob} \cite{prelim}.
In this computation we have taken the strange quark mass to be
$m_s=T_c$ and the two light quark masses to be degenerate,
$m_{ud}=0.03T_c$, such that it reproduces the correct value of
$m_\pi/m_\rho$. As can be seen from the figure, the value of the
ratio at $T_c$ is $\lambda_s\approx0.4$, in agreement with the value
of the Wroblewski parameter extracted from experiments, when the
freeze-out temperature is close to $T_c$ \cite{cleymans}.  It is
also a pleasant fact that at lower temperatures the ratio keeps
decreasing.

The dependence of this ratio on the valence quark masses was
investigated in \cite{rajarshi}, where it was shown that, in the
continuum limit, there was no dependence on valence quark mass
except near $T_c$. In the vicinity of $T_c$, and immediately below,
we found $\chi_s$ to be strongly dependent on $m_s$. It increases
as a function of $T/T_c$ and at large enough $T$ reaches the same
value as $\chi_u$, but it does this slowly when $m_s/T_c$ is large,
and faster when $m_s\ll T_c$. If the plasma contains strange quark
quasi-particles, as we argue later, then this behaviour could be a
kinematic effect, which measures the phase space for a thermal gluon
to split into a strange quark-antiquark pair. That the first effect
is dynamical and the second kinematical can be motivated by a study
of quantities which vanish in the SU(3) flavour symmetric limit.

\subsubsection{Flavour symmetry breaking}

\begin{figure}
\begin{center}
   \scalebox{0.5}{\includegraphics{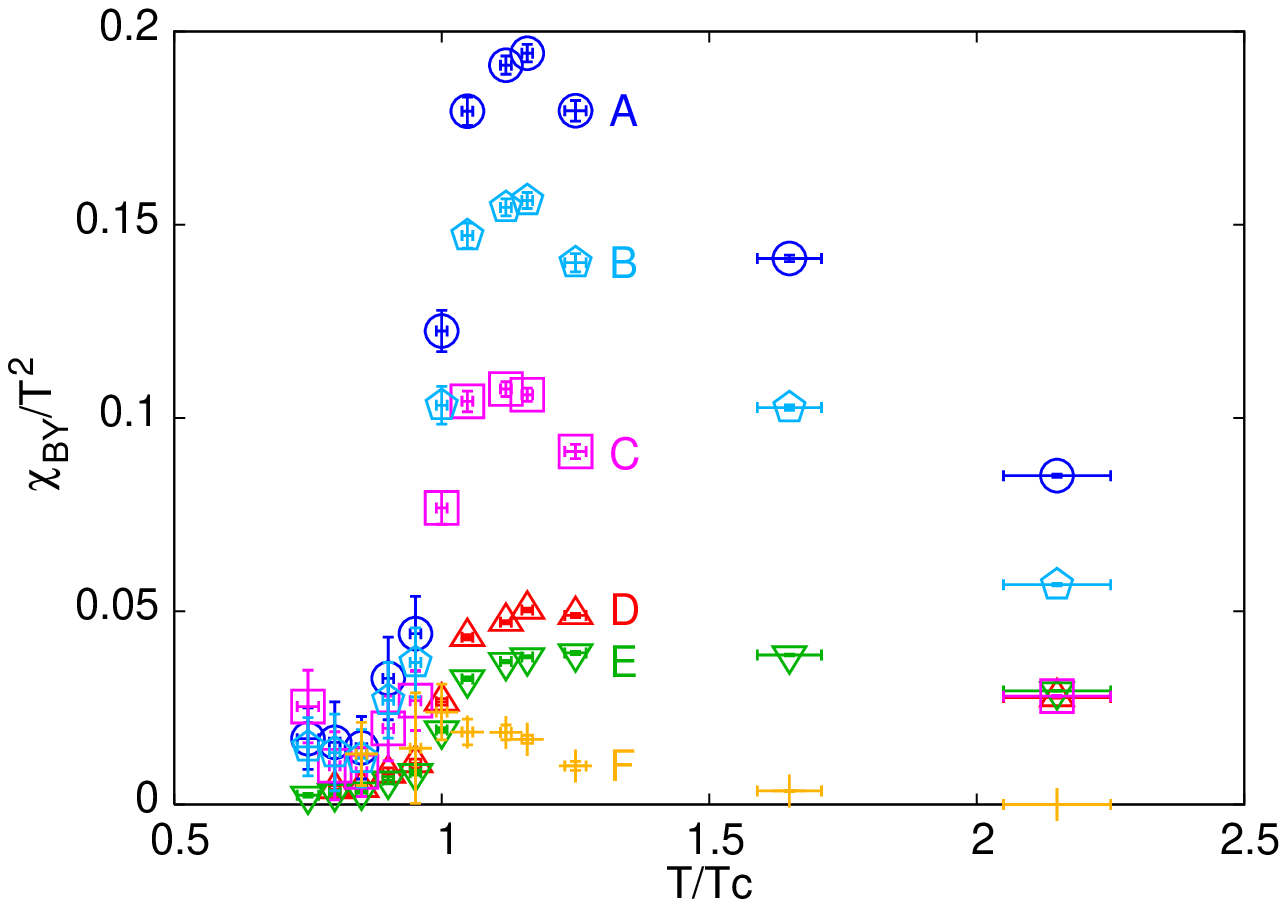}}
   \scalebox{0.5}{\includegraphics{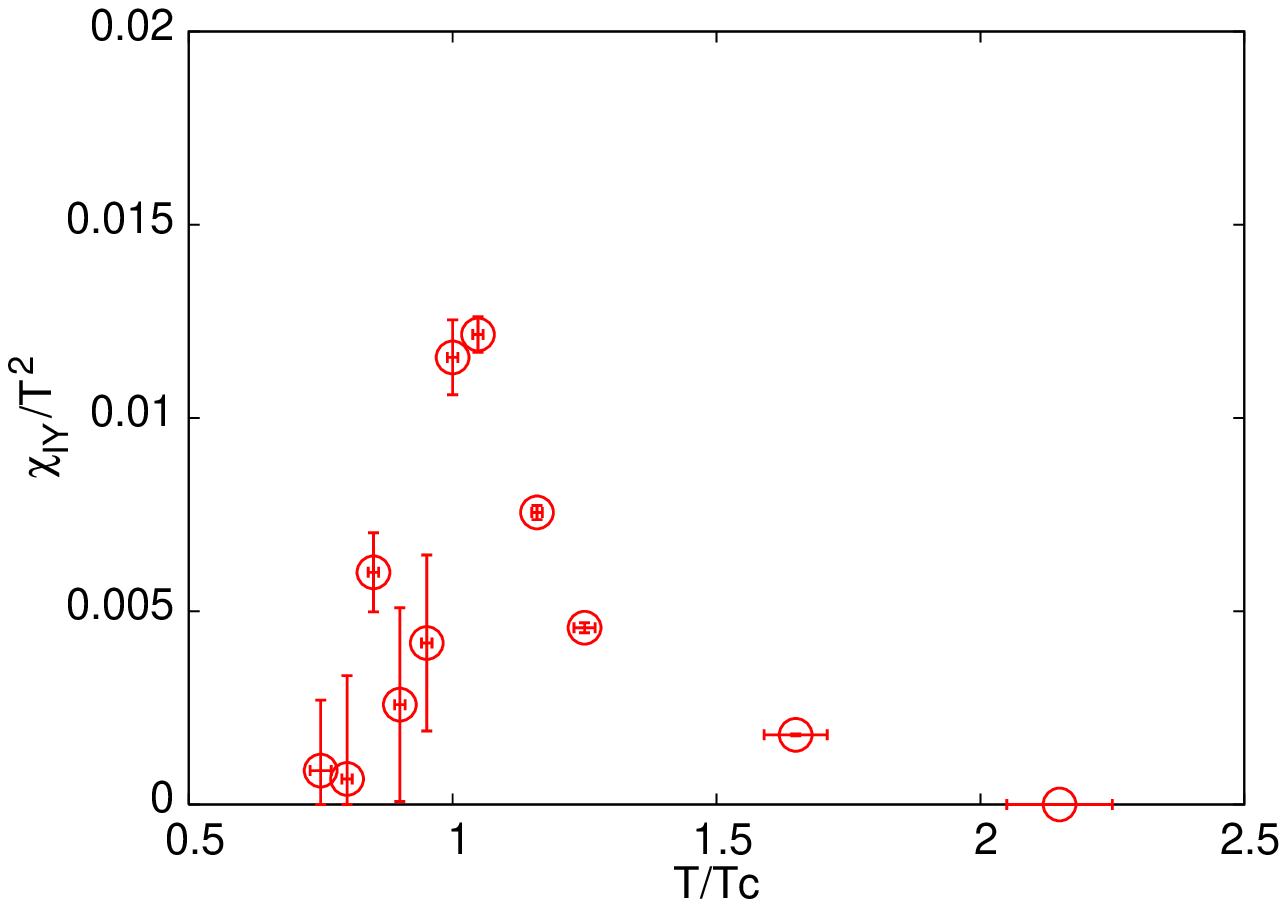}}
\end{center}
\caption{The first panel shows $\chi_{BY}/T^2$ as a function of $T/T_c$ for
   various patterns of SU(3) flavour symmetry breaking. Holding 
   $m_{ud}=0.1T_c$ constant we vary $m_s$ in (A) $m_s=T_c$, (B) $m_s=0.75T_c$,
   and (C) $m_s=0.5T_c$. Holding $\Delta_{us}=0.25T_c$ constant, we vary all
   the quark masses in (D) $m_s=0.75$ and (E) $m_s=T_c$. In (F)
   all the quark masses are small $m_{ud}=0.01T_c$, $m_s=0.1T_c$.
   The second panel shows $\chi_{IY}/T^2$, as a function of $T/T_c$ when
   $m_u=0.03T_c$, $m_d=0.1T_c$ and $m_s=T_c$.}
\label{fg.peak}\end{figure}

\begin{figure}
\begin{center}
   \scalebox{0.7}{\includegraphics{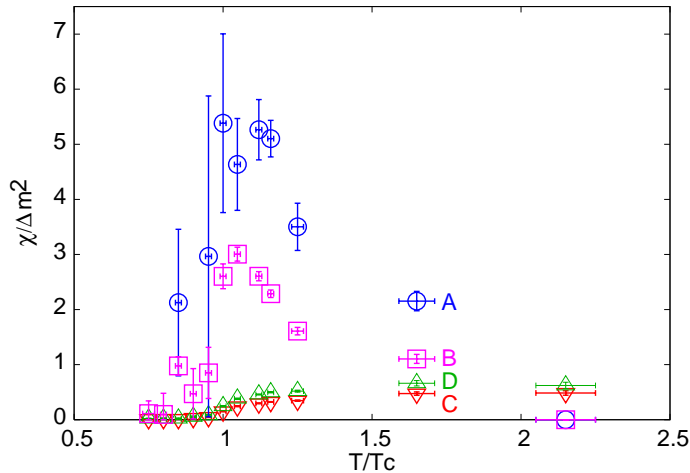}}
\end{center}
\caption{The flavour symmetry breaking matrix elements (A) $\chi_{BY}/\Delta_{us}^2$
   extracted with $m_{ud}=0.03T_c$ and $m_s=0.1T_c$, and (B) $A_{IY}=
   \chi_{IY}/\Delta_{ud}^2$  extracted using $m_u=0.03T_c$, $m_d=0.1T_c$
   and $m_s=T_c$, as a function of $T/T_c$. The kinematic suppression for
   realistic strange quark masses is clear from the significantly smaller
   values of $\chi_{BY}/\Delta_{us}^2$ when (C) $m_s=T_c$ and (D) $m_s=0.75T_c$.}
\label{fg.coeff}\end{figure}

Two off-diagonal susceptibilities show an interesting pattern---
$\chi_{BQ}$ and $\chi_{BY}$ are both continuous through $T_c$, but
peak in the vicinity of $T_c$. Since $\chi_{BQ}=\chi_{BY}/2$, as seen
from eqs.  (\ref{qnsa}) and (\ref{qnsb}), we show only the latter in 
Figure (\ref{fg.peak}) for various
values of quark masses explained in the caption. The figure also displays
$\chi_{IY}$ for $m_u\ne m_d$. Direct computations also show that
$\chi_{BQ}=\chi_{BY}=0$ when all three quark masses are equal, and that
$\chi_{IY}=0$ in the SU(2) symmetric limit--- providing an explicit
demonstration that non-zero values of these quantities are due to flavour
symmetry breaking (see the discussion in Appendix \ref{sc.qns}).

From a comparison of the cases (D) and (E) in Figure \ref{fg.peak}
it is clear that $\chi_{BY}$ is not only a function of
$\Delta_{us}=m_s-m_u$ and $T$ for large values of this asymmetry,
since the two curves are not coincident although they have equal
$\Delta_{us}$.  A careful look at the cases (A), (B) and (C) in the
same figure shows that when $m_s$ is comparable to $T_c$ then both
the position and the value of the peak in these QNS are dependent
on $m_s$. Explicit dependence of the flavour symmetry breaking
matrix elements on the actual value of $m_s$ (and not just the asymmetry
parameter) can only come as a kinematic effect. We try to confirm
the magnitude of this effect next.

In Figure \ref{fg.coeff} we display the values of the dimensionless
quantities $A_{IY}=\chi_{IY}/\Delta_{ud}^2$ and
$A_{BY}=\chi_{BY}/\Delta_{us}^2$, extracted using the computations
in which $\Delta_{us}$ and $\Delta_{ud}$ are much smaller than
$T_c$. It would be interesting to check the temperature range in
which these dimensionless quantities are computable in weak coupling
theory. In the same figure we also show $\chi_{BY}/\Delta_{us}^2$
when $\Delta_{us}$ is comparable to $T_c$. Its strong suppression
relative to the former case shows the kinematic effect which is responsible
for the shape of $C_{s/u}$ shown in Figure \ref{fg.wrob}.

The physics of the region just above $T_c$ is known to be complicated
when observed through gluonic variables such as $\Delta/T^4 =
(\epsilon-3P)/T^4$ (where $\epsilon$ is the energy density and $P$
the pressure) as well as the ratio of the lowest lying
screening masses in the CP-even and CP-odd sectors \cite{saumen}.
The peaks in $A_{BY}$ and $A_{IY}$ are the first observations of
interesting structures near $T_c$ in fermionic variables unconnected
with the order parameter. It would be interesting to see what
temperature range in this is explainable by weak coupling theory.

\subsection{Flavour carrying degrees of freedom}

\begin{figure}
\begin{center}
   \scalebox{0.7}{\includegraphics{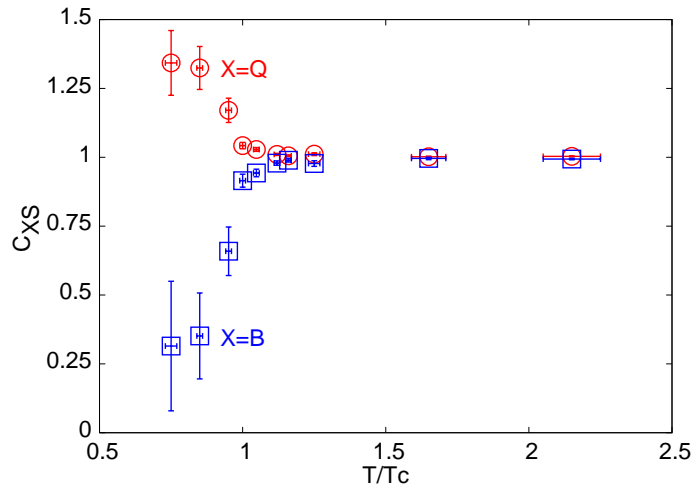}}
\end{center}
\caption{The robust variables $C_{BS}$ and $C_{QS}$, as functions of $T/T_c$.
   The quark masses used are $m_{ud}=0.1T_c$ and $m_s=T_c$, although in the
   high temperature phase there is no statistically significant dependence
   on the quark masses.}
\label{fg.bcs}\end{figure}

The question of which are the thermodynamically relevant degrees
of freedom in the QCD plasma is easier to answer in the quark sector
than in the gluon sector. The reason is that the multitude of flavour
quantum numbers allow us to look for ``linkage'' of flavour, \ie,
exciting one quantum number and seeing the magnitude of another
quantum number that is simultaneously excited.

\subsubsection{Strangeness carriers}

Robust variables involving off-diagonal QNS
serve precisely this purpose. In \cite{koch} the robust variable
\beq
   C_{BS} = -3C_{(BS)/S} = -3\,\frac{\chi_{BS}}{\chi_S} 
     = 1 + \frac{\chi_{us}+\chi_{ds}}{\chi_s}
     = 1 + C_{(us)/s}+C_{(ds)/s} = 1 + 2 C_{(us)/s}
\label{corr}\eeq
is identified as one which can distinguish between bound state QCD \cite{bqcd}
and the usual picture of the excitations in the plasma phase of
QCD (in the last expression above we have used eq. (\ref{qnsc})
and flavour SU(2) symmetry to write $C_{(us)/s}=C_{(ds)/s}$).  This is 
expected to have a value of unity if strangeness is carried by quarks (\ie,
$S=1$ always comes linked with $B=-1/3$).  In \cite{koch} it
was shown that bound state QGP gives a value of $C_{BS}\approx2/3$
(for $T>T_c$).

We present the first estimate for this quantity from lattice QCD
in Figure \ref{fg.bcs}.  In the low-temperature phase $C_{BS}$ is
very different from unity, but immediately above $T_c$ the value
is clamped to unity.  There is no statistically significant change
in $C_{BS}$ as $m_s/T_c$ is varied between 0.1 and 1.  Since the
statistical error bars are extremely small for $T\ge T_c$, this is
a strong statement which contrasts with the $m_s$ dependence of
$\lambda_s$ and $\chi_{BY}$.

Another interesting measure is the correlation of charge and
strangeness measured by the robust observable
\beq
   C_{QS} = 3C_{(QS)/S} = 1 - \frac{2\chi_{us}-\chi_{ds}}{\chi_s}.
\label{dorr}\eeq
When strangeness is carried by quarks one would expect this to be
unity (since $S=1$ comes with $Q=1/3$).  In Figure \ref{fg.bcs} we
have also shown the first measurement of $C_{QS}$.  Immediately above
$T_c$ it reaches close to unity with small errors.  As a result,
these two measurements together quite strongly indicate that unit
strangeness is carried by objects with baryon number $-1/3$ and
charge $+1/3$ in the high temperature phase of QCD, immediately
above $T_c$. 

Furthermore, eqs.\ (\ref{corr}, \ref{dorr}) indicate that our
observations imply that $\chi_{us}=0$, and hence strangeness carrying
excitations do not carry u or d flavour.  This is the most direct
lattice evidence to date that strangeness is linked to other
quantum numbers exactly as it would be for strange quarks, in the
high temperature phase of QCD; and that these linkages
are quite different below $T_c$.  Later in this section we show
that one should think of these as quasi-particles, dressed by the
strong residual interactions, rather than as elementary quarks.

Apart from the direct evidence of linkage between quantum numbers, we
also draw attention to the cryptic evidence in the temperature and
$m_s$ dependence of $C_{BS}$ and $C_{QS}$. The rapid change of $C_{BS}$
with $T$ (for $T<T_c$) has a natural explanation if the thermodynamics
is controlled by a spectrum of strange baryons such that the amount of
(anti-) strangeness per baryon increases with mass, and the masses are
larger than $T$. The temperature independence of the two quantities
above $T_c$ similarly implies that there is one excitation, which has
mass less than $T_c$. The fact that the values of these quantities do
not depend on $m_s$ within errors, for $T>T_c$, further implies that
the effective masses of these quasi-particles is less than $T_c$, so
that the infrared cutoff on the Dirac operator spectrum is provided by
$T$.  The next heavier quark, charm quark, does not affect the
thermodynamics of the QCD plasma in the range of temperatures we
investigate, since its mass is well beyond $T$.  However, these heavier
quarks do probe changes in other aspects of physics, such as screening,
as is evident from \cite{jpsi}.

\subsubsection{The light quark sector}

\begin{figure}[bht]
\begin{center}
   \scalebox{0.7}{\includegraphics{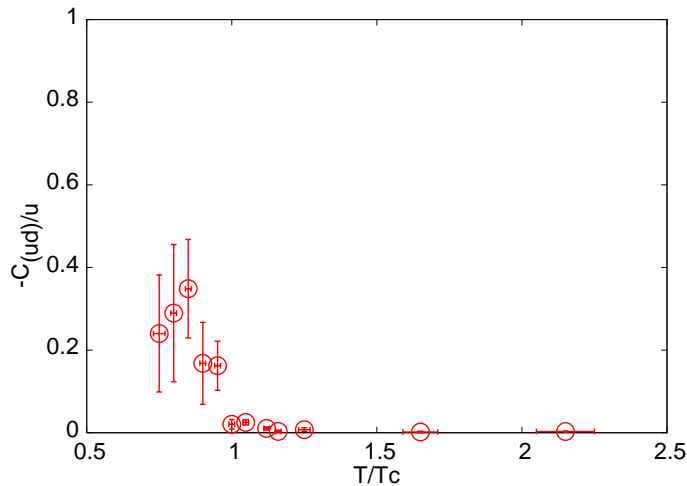}}
\end{center}
\caption{The robust variable $-C_{(ud)/u}$ which measures the correlation
   between u and d flavours for $m_{ud}=0.1T_c$. It is positive in the low temperature phase
   since u quarks are found along with d antiquarks in charged pions,
   and vanishes in the high temperature phase, indicating that u and d
   are fully decorrelated in the plasma. There is no statistically
   significant dependence on $m_{ud}$ in the high temeperature phase.}
\label{fg.lightqk}\end{figure}

In transplanting these methods to the light
quark sector, we find that the composite QNS, $\chi_{BI}$ and
$\chi_{QI}$, are not informative, since the quark of one flavour
has the same isospin as the antiquark of the other flavour. One way
to extract information on the degrees of freedom would be to consider
QNS of G-parity. However, it is more transparent to turn to the
flavoured QNS $\chi_{ud}\propto\langle\N_u\N_d\rangle$. We can then use
the quantity
\beq
   C_{(ud)/u} = \frac{\chi_{ud}}{\chi_u},
\label{light}\eeq
which looks at the linkage between u and d flavours in the same
way that $C_{(QS)/S}$ looked for linkage of strangeness
and charge. Our results are plotted in the first panel of Figure
\ref{fg.lightqk}. In the hadronic phase it
is non-vanishing because of charged pions, and
negative because in these mesons each u comes with a $\overline{\rm
d}$, and vice versa. In the QGP phase the vanishing of this normalized
covariance implies that a particle with $u$ quantum number does not
exhibit $d$ quantum numbers. 

Further tests come from investigating
\beqa
\nonumber
   C_{(BU)/U} &=& C_{(BD)/D} = \frac13(1+C_{(ud)/u}+C_{(us)/u}),\\
\nonumber
   C_{(QU)/U} &=& \frac13(2-C_{(ud)/u}-C_{(us)/u}),\\
   C_{(QD)/D} &=& -\frac13(1-2C_{(ud)/u}+C_{(us)/u}).
\label{crosscorr}\eeqa
The vanishingly small values of $C_{(ud)/u}$ and $C_{(us)/u}$ imply that
the u flavour is carried by excitations with baryon number $+1/3$
and charge $+2/3$, whereas the d flavour is carried by particles with
baryon number $+1/3$ and charge $-1/3$. These are, therefore, quark
quasi-particles.

\subsubsection{Quasi-quarks}

One might wonder why we talk of, for example, baryon number 1/3
when the measurements even at $2T_c$ differ from this number by a
few parts in a thousand. What does this small but statistically
significant deviation tell us? The answer is that it says something
about the spatial structure of the quasi-particle.  If flavour were
carried by pointlike bare quarks, then $\chi_{ud}$ and $\chi_{us}$
would be precisely zero.  However, interactions dress each quark
into a spatially extended quasi-particle, and a thermodynamic average
probes the spatial dimension of the charge with a resolution of
$1/2\pi T$.  When $T$ is sufficiently large, so that the gauge
coupling is sufficiently small, this structure can be computed in
weak coupling theory. As the coupling grows, the perturbative
computation fails quantitatively, but as long as the correction to
the charge or baryon number remains small, one can fruitfully talk
of quasi-quarks.

\begin{figure}[bht]
\begin{center}
   \scalebox{0.7}{\includegraphics{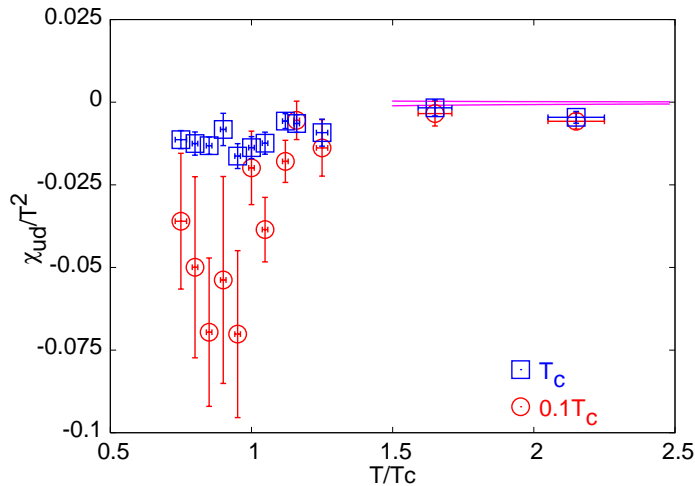}}
\end{center}
\caption{The off-diagonal QNS $\chi_{ud}/T^2$ for two different quark
   masses compared to weak coupling theory. The band includes uncertainties
   due to neglected of higher loop effects, the effect of changing from
   one-loop to two-loop computation of the running coupling, and 
   statistical uncertainties in the determination of 
   $T_c/\Lambda_{\overline{\scriptscriptstyle MS}}$.}
\label{fg.chiud}\end{figure}

In Figure \ref{fg.chiud} we show the flavour off-diagonal QNS $\chi_{ud}/T^2$
for two different quark masses, along with the prediction of weak coupling
perturbation theory \cite{bir}---
\beq
   \frac{\chi_{ud}}{T^2} = -\frac{10}{27\pi^3}\alpha_s^3
         \log\left(\frac c{\alpha_s}\right),
\label{bir}\eeq
where $c$ is a constant whose evaluation requires larger number of
loops in the perturbation theory. The strong coupling, $\alpha_s$,
has been evaluated to two loop accuracy at scale $2\pi T$ with the
estimate $T_c/\Lambda_{\overline{\scriptscriptstyle MS}}=0.49\pm0.05$
\cite{precise}.  This variation in
$T_c/\Lambda_{\overline{\scriptscriptstyle MS}}$, a variation of
$c$ by two orders of magnitude, $0.1\le c\le10$, and the variation
in $\alpha_s$ in going from one loop to the two loop expression are
included in the band in the figure. We find that in this range of
temperature the prediction is somewhat smaller than the lattice
data. Since this is not a robust variable, it is possible that
taking the continuum limit will improve the agreement between the
two. However, it is clear that as one comes closer to $T_c$ the
disagreement increases, although the magnitude of $C_{(ud)/u}$
remains small. Thus, it seems that a Fermi gas picture, which may
be valid at large $T/T_c$, gives way to something more complicated
as one approaches $T_c$, although the quantum numbers are linked
in exactly the same way as for the elementary quarks. This is the
meaning of quasi-quarks.

\section{Summary}

We have presented an extensive computation of many different quark
number susceptibilities (see Appendix \ref{sc.qns} for the definitions).
All the diagonal QNS, and some of the off-diagonal QNS, track the
phase structure of QCD--- being small in the confined phase and
crossing over to larger values in the high temperature phase of
QCD, as shown in Figure \ref{fg.qns}.

An important observation was that ratios of QNS, $C_{A/B}$, defined
in eq.\ (\ref{ratio}), are robust variables which depend weakly on
the lattice spacing and the sea quark content of QCD in the high
temperature phase, as shown in Figure \ref{fg.ratio}. These ratios
can be compared to experimentally determined ratios of variances
(or covariances) in event-to-event fluctuations of conserved quantum
numbers. The relative magnitudes of the diagonal QNS are among these
robust observables, and we found the ordering $\chi_S > \chi_Q > \chi_Y
> \chi_I > \chi_B$, shown in Figure \ref{fg.robust}.

A second set of results concerns the thermal production rate of
strange quarks. It has been argued \cite{conti} that under certain
(testable) conditions the Wroblewski parameter is the robust observable
$C_{s/u}$. While it is insensitive to the sea quark
content of QCD, it is known to depend sensitively on the valence
quark masses \cite{rajarshi}. Here we have determined this quantity
for realistic values of the strange and light quark masses (see
Figure \ref{fg.wrob}). 

We attributed this dependence on $m_s$ to kinematic effects visible
when $m_s\simeq{\cal O}(T_c)$. However, kinematic effects should
manifest themselves in other quantities as well. We tested this hypothesis by examining
certain QNS which vanish in the flavour symmetric limit. We extracted
the matrix elements which are quadratic in the flavour symmetry
breaking mass differences, $\Delta_{us}$, when $\Delta_{us}\ll T_c$.
By comparing these (in Figure \ref{fg.coeff}) to the corresponding
quantities when $\Delta_{us}\approx{\cal O}(T_c)$, we demonstrated
the presence of such kinematic effects in other quantities as well.
The flavour symmetry breaking matrix elements themselves (Figure \ref{fg.coeff})
peak at $T$ slighly larger than $T_c$, and are the first known example
of observables in the quark sector of QCD which parallel similiar
structures seen in the gluon sector.

Our final result is that the high temperature phase of QCD essentially
consists of quasi-quarks. We demonstrated this by observing that
unit strangeness is carried by something which has baryon number
$-1/3$ and charge $1/3$, as shown in Figure \ref{fg.bcs}. Part of
the argument is that this correlation does not depend on the strange
quark mass even when it is as large as $T_c$. Similarly, in the
light quark sector one finds that u and d quantum numbers are not
produced together (Figure \ref{fg.lightqk}). Through eqs.\
(\ref{crosscorr}) we found that this implies that the u flavour is
carried by excitations with baryon number $+1/3$ and charge $+2/3$,
whereas the d flavour is carried by particles with baryon number
$+1/3$ and charge $-1/3$.

We presented an argument that the carriers of these quantum numbers
are not elementary quarks but their dressed counterparts which are
called quasi-quarks. This argument involved the comparison of
$\chi_{ud}$ with a weak-coupling prediction, which is shown in
Figure \ref{fg.chiud}. The key point is that this comparison fails
badly as one approaches $T_c$, although the correlations of flavour
quantum numbers remains as they would for quarks. A similar
comparison of weak coupling prediction with lattice results for the
diagonal QNS $\chi_u$ also fails near $T_c$, leading us to the same
conclusion.

The argument about the existence of quasi-quarks in the high
temperature phase of QCD depends on the examination of robust
variables given in eq.\ (\ref{example}). It is useful to note that
their use is not restricted to the lattice. It is also possible to
measure them in heavy-ion collisions and thereby deduce the nature
of excitations in the fireball produced in these collisions.

We end by pointing out that we have not studied the low-temperature
phase of QCD in much detail here. This is an interesting problem,
which we have touched upon very briefly in the discussion of $C_{BS}$
and $C_{QS}$, and has been left for the future.

{\bf Acknowledgements}: This computation was carried out on the
Indian Lattice Gauge Theory Initiative's CRAY X1 at the Tata Institute
of Fundamental Research. It is a pleasure to thank Ajay Salve for
his administrative support on the Cray. Part of this work was done
during a visit under Indo-French (IFCPAR) project, 3104-3 to SPhT,
Saclay. The hospitality of Saclay and IFCPAR's support is gratefully
acknowledged.

\appendix
\section{Susceptibilities in equivalent ensembles}\label{sc.qns}

Corresponding to every conserved charge, $\Q$, under the global
symmetries of a theory, one can introduce a chemical potential,
$\mu$, into thermodynamics, by adding to the action a source term
$\mu\Q$. In QCD, at finite quark mass, one has $SU(N_f)$ vector
flavour symmetry. Corresponding to each of the $N_f$ flavours, one
can introduce a chemical potential $\mu_f$ ($f=u$, $d$, $s$, \etc)
through the term
\beq
   J = \sum_f \mu_f \N_f = \mu^T \N,
\label{source}\eeq
where $\N_f$ is the number operator for quarks of flavour $f$, whose
expectation value is the number of quarks minus the number of
antiquarks. The last expression just rewrites the sum as a dot
product of the vector of intensive variables $\mu$ with the vector
of extensive variables $\N$. This corresponds to a grand canonical
ensemble in which the chemical potential on each of the quark
flavours can be tuned independently. In the corresponding canonical
ensemble the number of quarks of each flavour is kept fixed.

The number densities, which are the first derivative of the pressure
with respect to the chemical potential, and the quark number
susceptibilities (QNS), which are the second derivatives, have been
defined before \cite{milc,conti}.  Here we use the notation of
\cite{conti} for the QNS. We shall also use a higher order
susceptibility, for which we use the notation of \cite{endpt}.

It is usually more convenient to define chemical potentials for
variables which are easier to control in experiments--- such as the
baryon number, $B$, the electric charge, $Q$, the third component
of isospin, $I_3$, or the hypercharge, $Y$. Any choice of $N_f$
variables corresponds to a different choice of ensemble to describe
the same physics. The description in terms of flavours given above
can then be translated into the new ensemble by a simple linear
transformation
\beq
   J = \mu^T M^{-1} M\N = (\mu')^T \N', \qquad i.e.,\qquad \N'=M\N
     \quad{\rm and}\quad \mu'=(M^{-1})^T\mu=(M^T)^{-1}\mu.
\label{transf}\eeq
Clearly, the partition function being the same, the physics remains
invariant under these redefinitions. The choice of a given $M$
corresponds to putting coordinates in Gibbs space.

One is usually interested in thermodynamics quantities or response
functions which are obtained by taking derivatives of the free
energy or pressure with respect to the chemical potentials. Note
that by the definitions in eq.\ (\ref{transf}), one has $\mu=M^T\mu'$.
The chain rule for differentiation then tells us that
\beq
   \frac{\partial}{\partial\mu'_i} = 
   \frac{\partial\mu_j}{\partial\mu'_i} \frac{\partial}{\partial\mu_j}
   = (M^T)_{ji} \frac{\partial}{\partial\mu_j}
   = M_{ij} \frac{\partial}{\partial\mu_j}
\label{pder}\eeq
The fact that this gives back the original definitions of the $\N'$s
in terms of the $\N$s is a consistency check of the formalism. We
illustrate the uses of this formalism for the cases of $N_f=2$ and
$N_f=3$ below.

\subsection{$N_f=2$}

\subsubsection{The $B$, $I_3$ ensemble}

In the two flavour case, one can transform from the flavour basis to the set
\beqa
\nonumber
   B=\frac13(\N_u+\N_d),\qquad&&\mu_B=\frac32(\mu_u+\mu_d),\\
   I_3=\frac12(\N_u-\N_d),\qquad&&\mu_I=\mu_u-\mu_d.
\label{twoflava}\eeqa
Inverting the relation between chemical potentials one obtains
$\mu_u=\mu_B/3+\mu_I/2$ and $\mu_d=\mu_B/3-\mu_I/2$.
In the ensemble where $\mu_I=0$, one then gets $\mu_u=\mu_d=\mu_B/3$. The
number densities are given by eq.\ (\ref{twoflava}). The quark number
susceptibilities are---
\beqa
\nonumber
   \chi_B &=& \frac19(\chi_{uu}+\chi_{dd}+2\chi_{ud}) = \frac29(\chi_u+\chi_{ud}),\\
\nonumber
   \chi_I &=& \frac14(\chi_{uu}+\chi_{dd}-2\chi_{ud}) = \frac12(\chi_u-\chi_{ud}),\\
   \chi_{BI} &=& \frac16(\chi_{uu}-\chi_{dd}) = 0,
\label{qnstwoa}\eeqa
where the first expression in each case is the most general, and the second is
obtained for exact vector SU(2) flavour symmetry $m_{ud}$. If this symmetry is
broken then $\chi_{BI}$ should become non-zero. For small values of the
symmetry breaking parameter $\Delta_{ud}=m_d-m_u$ (we will follow the 
convention that $m_u\le m_d\le m_s$),
\beq
   \chi_{BI} = A_{BI} \Delta_{ud}^2, \qquad{\rm which\ yields}\qquad
   \frac{\chi_{BI}}{T_c^2} = A_{BI} \left(\frac{\Delta_{ud}}{T_c}\right)^2,
\label{isobrk}\eeq
where $A_{BI}$ is a dimensionless number. We present results for this quantity 
in Figure \ref{fg.coeff}. In the low temperature phase we expect 
that $A_{BI}$ is a
non-perturbative quantity, but that it should be computable in chiral
perturbation theory. It would be interesting to check how far the weak coupling
theory in the high temperature phase agrees with our determination of $A_{BI}$.

\subsubsection{The $B$, $Q$ ensemble}

One may choose to work in another ensemble given by
\beqa
\nonumber
   B=\frac13(\N_u+\N_d),\qquad&&\mu_B=\mu_u+2\mu_d,\\
\nonumber
   Q=\frac13(2\N_u-\N_d),\qquad&&\mu_Q=\mu_u-\mu_d.
\label{twoflavb}\eeqa
For $\mu_Q=0$, one gets again the expected result $\mu_u=\mu_d=\mu_B/3$.
The quark number susceptibilities in this ensemble are---
\beqa
\nonumber
   \chi_B &=& \frac19(\chi_{uu}+\chi_{dd}+2\chi_{ud}) = \frac29(\chi_u+\chi_{ud}),\\
\nonumber
   \chi_Q &=& \frac19(4\chi_{uu}+\chi_{dd}-4\chi_{ud}) = \frac19(5\chi_u-4\chi_{ud}),\\
   \chi_{BQ} &=& \frac19(2\chi_{uu}-\chi_{dd}+\chi_{ud}) = \frac19(\chi_u+\chi_{ud}),
\label{qnstwob}\eeqa
where the last expression in each line holds in the special case of $m_{ud}$. Note
that $\chi_{BB}$ is the same in both the ensembles. This follows from the fact that
the definition of the baryon number is the same.

\subsection{$N_f=3$}

\subsubsection{The $B$, $I_3$, $Y$ ensemble}

The variables in this ensemble are
\beqa
\nonumber
   B=\frac13(\N_u+\N_d+\N_s),\qquad&& \mu_B=\mu_u+\mu_d+\mu_s,\\
\nonumber
   I_3=\frac12(\N_u-\N_d),\qquad&& \mu_I=\mu_u-\mu_d,\\
   Y=\frac13(\N_u+\N_d-2\N_s)&& \mu_Y=\frac12(\mu_u+\mu_d-2\mu_s).
\label{defseta}\eeqa
The six independent quark number susceptibilities are
\beqa
\nonumber
   \chi_B &=& \frac19(\chi_u+\chi_d+\chi_s+2\chi_{ud}+2\chi_{us}+2\chi_{ds}) 
              =  \frac19(2\chi_u+\chi_s+2\chi_{ud}+4\chi_{us}),\\
\nonumber
   \chi_I &=& \frac14(\chi_u+\chi_d-2\chi_{ud}) = \frac12(\chi_u-\chi_{ud}),\\
\nonumber
   \chi_Y &=& \frac19(\chi_u+\chi_d+4\chi_s+2\chi_{ud}-4\chi_{us}-4\chi_{ds}) 
              =  \frac29(\chi_u+2\chi_s+\chi_{ud}-4\chi_{us}),\\
\nonumber
   \chi_{BI} &=& \frac16(\chi_u-\chi_d+\chi_{us}-\chi_{ds}) = 0,\\
\nonumber
   \chi_{BY} &=& \frac19(\chi_u+\chi_d-2\chi_s+2\chi_{ud}-\chi_{us}-\chi_{ds}) 
              =  \frac29(\chi_u-\chi_s+\chi_{ud}-\chi_{us}),\\
   \chi_{IY} &=& \frac19(\chi_u-\chi_d-2\chi_{us}+2\chi_{ds}) =0.
\label{qnsa}\eeqa
As before, the last set of expressions on each line holds only for
$m_{ud}\ne m_s$.  Similiar to eq.\ (\ref{isobrk}), one can define
$A_{BI}$ here, and also $A_{IY}=\chi_{IY}/\Delta_{ud}^2$, both of
which are generally non-vanishing when SU(2) flavour symmetry is broken.

In the SU(3) symmetric limit, $m_u = m_d = m_s$, the three off-diagonal
susceptibilities vanish, \ie, $\chi_{BI} = \chi_{BY} = \chi_{IY} =
0$.  Also, in this limit one has $2\chi_I = 3\chi_Y/2 = \chi_u-\chi_{ud}$
and $3\chi_B = \chi_u+2\chi_{ud}$. In the low temperature phase the
breaking of vector SU(3) symmetry produces the mass difference
between the pion and the K meson. At sufficiently high temperature,
when the strange quark mass is less than the Matsubara frequency,
$m_s<2\pi T$, the theory becomes effectively SU(3) symmetric, and
the above relations should hold. In the high temperature phase of
QCD one also has $\chi_{ud}\propto g^5\log g\to0$ \cite{bir}, so
one should obtain $2\chi_I = 3\chi_Y/2 = 3\chi_B$.

\subsubsection{The $B$, $Q$, $Y$ ensemble}

Another useful set of charges and associated chemical potentials is
\beqa
\nonumber
   B=\frac13(\N_u+\N_d+\N_s),\qquad&&\mu_B=\mu_u+\mu_d+\mu_s,\\
\nonumber
   Q=\frac13(2\N_u-\N_d-\N_s),\qquad&&\mu_Q=\mu_u-\mu_d,\\
   Y=\frac13(\N_u+\N_d-2\N_s),\qquad&&\mu_Y=\mu_d-\mu_s.
\label{defsetb}\eeqa
The three susceptibilities $\chi_B$, $\chi_Y$ and $\chi_{BY}$ are the same
as in the previous ensemble. The remaining susceptibilities are
\beqa
\nonumber
   \chi_Q &=& \frac19(4\chi_u+\chi_d+\chi_s-4\chi_{ud}-4\chi_{us}+2\chi_{ds}) 
              =  \frac19(5\chi_u+\chi_s-4\chi_{ud}-2\chi_{us}),\\
\nonumber
   \chi_{BQ} &=& \frac19(2\chi_u-\chi_d-\chi_s+\chi_{ud}+\chi_{us}-2\chi_{ds})
              =  \frac19(\chi_u-\chi_s+\chi_{ud}-\chi_{us}),\\
   \chi_{QY} &=& \frac19(2\chi_u-\chi_d+2\chi_s+\chi_{ud}-5\chi_{us}+\chi_{ds}) 
              =  \frac19(\chi_u+2\chi_s+\chi_{ud}-4\chi_{us}).
\label{qnsb}\eeqa
As before, the last set of expressions on each line holds only for
$m_{ud}\ne m_s$.  Note that in this limit one has $\chi_{BQ}=\chi_{BY}/2$.

In the SU(3) symmetric limit, $m_{ud}=m_s$, two of the off-diagonal
susceptibilities vanish, \ie, $\chi_{BQ}=\chi_{BY}=0$. Also, in
this limit one has $\chi_Q=\chi_Y=2\chi_{QY}=2(\chi_u-\chi_{ud})/3$
and $\chi_B=(\chi_u+2\chi_{ud})/3$. As before, in the high temperature
limit, when the quark masses are less than the temperature, the
theory becomes effectively SU(3) symmetric. Then taking $\chi_{ud}=0$,
one should obtain $\chi_Q=\chi_Y=2\chi_B=2\chi_{QY}$.

It is interesting to examine this in a theory of massless free fermions. The
free energy is given by
\beq
   F = N_c V \sum_{f=u,d,s} \left[\frac{7\pi^2}{180} T^4 +\frac16\mu^2 T^2
         +\frac1{12\pi^2}\mu^4\right].
\label{fft}\eeq
Substituting the values of the flavour chemical potential by the
appropriate combination of $\mu_B$, $\mu_Q$ and $\mu_Y$, and taking
the derivatives, we find that $\chi_{BY} = \chi_{BQ} = 0$. Also,
$\chi_Q = \chi_Y = 2\chi_B = 2\chi_{QY} = 2N_c/9$.  For massive
free fermions, when $T$ is much larger than the fermion mass, the
same results would hold.

When SU(3) symmetry is broken through the parameter $\Delta_{us}=m_s-m_u$
(we assume that SU(2) symmetry still holds) then for small $\Delta_{us}$
one may again write
\beq
   \chi_{BY}=A_{BY}\Delta_{us}^2 \qquad{\rm and}\qquad
   \chi_{BQ}=A_{BQ}\Delta_{us}^2.
\label{flvbrk}\eeq
As before, we expect $A_{BY}$ and $A_{BQ}$ to be non-perturbative but
computable in chiral perturbation theory in the low temperature phase.
It would be interesting to compare our results (presented later) with
weak coupling theory in the high temperature phase.

\subsubsection{The $B$, $Q$, $S$ ensemble}

From the experimental point of view, it may be interesting to use the set
\beqa
\nonumber
   B=\frac13(\N_u+\N_d+\N_s),\qquad&&\mu_B=\mu_u+2\mu_d,\\
\nonumber
   Q=\frac13(2\N_u-\N_d-\N_s),\qquad&&\mu_Q=\mu_u-\mu_d,\\
   S=-\N_s,\qquad&&\mu_S=\mu_d-\mu_s.
\label{defsetc}\eeqa
Note that we have used the standard convention where the strangeness of the
antistrange quark is $+1$. The three susceptibilities, $\chi_B$, $\chi_Q$
and $\chi_{BQ}$ are as before. The remainder are
\beqa
\nonumber
   \chi_S &=& \chi_s,\\
\nonumber
   \chi_{BS} &=& -\frac13(\chi_s+\chi_{us}+\chi_{ds}) = -\frac13(\chi_s+2\chi_{us}),\\
   \chi_{QS} &=& \frac13(\chi_s-2\chi_{us}+\chi_{ds}) =  \frac13(\chi_s-\chi_{us}).
\label{qnsc}\eeqa
As always, the last set of expressions on each line holds only for $m_{ud}\ne m_s$.

\subsubsection{The $B$, $Q$, $U$ ensemble}

For technical questions about the light quark sector it is useful to work
in the ensemble with
\beqa
\nonumber
   B&=&\frac13(\N_u+\N_d+\N_s),\\
\nonumber
   Q&=&\frac13(2\N_u-\N_d-\N_s),\\
   U&=&\N_u,
\label{defsetu}\eeqa
The three susceptibilities, $\chi_B$, $\chi_Q$ and $\chi_{BQ}$ are as before.
The rest are
\beqa
\nonumber
   \chi_U &=& \chi_u,\\
\nonumber
   \chi_{BU} &=& \frac13(\chi_u+\chi_{ud}+\chi_{us}),\\
   \chi_{QU} &=& \frac13(2\chi_u-\chi_{ud}-\chi_{us}).
\label{qnsu}\eeqa
Changing to an ensemble where $U$ is replaced by $D=\N_d$ changes the QNS to
\beqa
\nonumber
   \chi_D &=& \chi_u,\\
\nonumber
   \chi_{BD} &=& \frac13(\chi_u+\chi_{ud}+\chi_{us}),\\
   \chi_{QD} &=& -\frac13(\chi_u-2\chi_{ud}+\chi_{us}),
\label{qnsd}\eeqa
where we have used SU(2) symmetry. This gives $\chi_U=\chi_D$ and $\chi_{BU}=\chi_{BD}$.

\end{document}